\documentclass[superscriptaddress,preprintnumbers,amsmath,amssymb,pre,twocolumn]{revtex4-1}

\usepackage{mathrsfs}
\usepackage{graphicx}
\usepackage[perpage,symbol]{footmisc}
\usepackage{dcolumn}
\usepackage{bm}
\usepackage{bbm}
\usepackage{color}
\usepackage{esint}
\usepackage{lineno}

\begin{document}

\title{Characterizing Real-space Topology in Rice-Mele Model by Thermodynamics}

\author{Jia-Bin You}
\affiliation{State Key Laboratory of Magnetic Resonance and Atomic
and Molecular Physics, Wuhan Institute of Physics and Mathematics,
Chinese Academy of Sciences, Wuhan 430071, China}
\affiliation{Department of Electronics and Photonics, Institute of High Performance Computing, 1 Fusionopolis Way, 16-16 Connexis, Singapore 138632, Singapore}
\author{Wanli Yang}
\email{ywl@wipm.ac.cn}
\affiliation{State Key Laboratory of Magnetic Resonance and Atomic
and Molecular Physics, Wuhan Institute of Physics and Mathematics,
Chinese Academy of Sciences, Wuhan 430071, China}

\date{\today}

\begin{abstract}

The thermodynamic quantities which are related to energy-level statistics are used to characterize the real-space topology of the Rice-Mele model. Through studying the energy spectrum of the model under different boundary conditions, we found that the non-normalizable wave function for the infinite domain is reduced to the edge state adhered to the boundary. For the finite domain with symmetric boundary condition, the critical point for the topological phase transition is equal to the inverse of the domain length. In contrast, the critical point is zero for the semi-infinite domain. Additionally, the symmetry of the energy spectrum is found to be sensitive to the boundary conditions of the Rice-Mele model, and the emergence of the edge states as well as the topological phase transition can be reflected in the thermodynamic properties. A potentially practical scheme is proposed for simulating the Rice-Mele model and detecting the relevant thermodynamic quantities in the context of Bose-Einstein condensate.

\end{abstract}


\maketitle

\section{Introduction}

The investigation of quantum simulation in photonic systems has received much attention in the past decade \cite{Nature.429.277,NewJ.Phys.17.013018,PhysRevLett.110.260405,PhysRevLett.91.107902,PhysRevA.81.043609,Nat.Phys.1.23,Sowinski:15,PhysRevLett.110.076401,PhysRevLett.111.026802,PhysRevLett.109.233004,PhysRevA.90.063638,PhysRevB.90.195112,SR.6.21775,PhysRevA.89.052313,PhysRevA.92.041805,PhysRevA.93.033631,RevModPhys.89.011004}. In particular, there has been great interest in mimicking the topological phase transition with ultracold atom which provides a convenient controllable platform for studying condensed-matter physics via photonic processes. Recently, one of the simplest one-dimensional models with nontrivial topology, namely, the Rice-Mele model \cite{PhysRevLett.49.1455}, has been experimentally constructed with ultracold bosons in optical lattices \cite{Atala2013,Nat.Phys.12.296,Nature.422.147} and the corresponding Zak phase was measured. The Rice-Mele model originally arising from the study of conducting polymers \cite{RevModPhys.60.781,PhysRevLett.62.2747,PhysRevLett.42.1698,Jackiw1981253,PhysRevLett.47.986,PhysRevLett.49.1455,RevModPhys.83.1057,PhysRevB.87.054501,YOU2014189,J.Phys.Condens.Matter.27.225302,J.Stat.Mech.P10032} is particularly interesting owing to its unusual topological properties characterized by
a nontrivial Zak phase \cite{PhysRevLett.62.2747,PhysRevB.84.195452}, gauge-invariant cumulants and moments \cite{PhysRevA.95.062104,Yahyavi2014}, domain-wall solitons \cite{PhysRevLett.42.1698,PhysRevLett.88.180401}, and a fractional charge \cite{Jackiw1981253,PhysRevLett.47.986}.

In this work, the real-space topology of the Rice-Mele model is characterized by thermodynamics. Here, we study the Rice-Mele model from the thermodynamic aspect. We first consider the model in the infinite, semi-infinite, and finite domains to obtain the energy spectrum in different boundary conditions. It is found that the edge state arises from the non-normalizable state in the infinite domain. The Rice-Mele model itself satisfies time-reversal-mirror (TRM) symmetry which leads to the symmetric energy spectrum of the bulk, however, we find that the spectral symmetry is sensitive to the boundary conditions. Several thermodynamic quantities, such as the particle and energy fluctuations as well as the entropy, are used to describe the statistical properties of the model, and the differences of these thermodynamic quantities can be used to characterize the edge state as well as the topological property of the system. We find that when the semi-infinite domain is considered, the in-gap peak stemming from the edge state in the fluctuations and entropy profiles is asymmetric; however, the finite domain with symmetric boundary condition (SBC) leads to symmetric in-gap peaks in the thermodynamic quantities when the system is in the topological phase. Experimentally, several thermodynamic quantities, such as temperature, pressure, energy, entropy, position, and momentum distributions could be measurable and inferred from the density profile of the condensate, by using the time-of-flight technique and absorption imaging  in the cold atom experiments \cite{RevModPhys.80.885,Kinast1296,Ku563,Luo2009}.

The manuscript is organized as follows. In Sec. \ref{boundary_value_problem}, we calculate the energy spectrum in infinite, semi-infinite, and finite domains to study the real-space topology in the Rice-Mele model. In Sec. \ref{thermodynamic_properties}, we use the thermodynamic quantities to further describe the edge state and topology in the model. In Sec. \ref{exp_realization}, we discuss the experimental realization and thermodynamic measurements of the Rice-Mele model in the cold atom setup. In Sec. \ref{conclusion}, we conclude that the thermodynamics can be used to characterize the real-space topology of the Rice-Mele model.

\section{ENERGY SPECTRUM AND REAL-SPACE TOPOLOGY IN DIFFERENT BOUNDARY CONDITIONS}
\label{boundary_value_problem}

The Hamiltonian of the Rice-Mele model can be expressed as $H=\sigma_{z}\Delta+\sigma_{x}\delta{J}+i\sigma_{y}J\partial_{x}$, which could be reduced to the celebrated Su-Schrieffer-Heeger model \cite{PhysRevLett.42.1698} with chrial symmetry $\sigma_{z}H\sigma_{z}=-H$ if $\Delta=0$. Although the chiral symmetry is broken in the general case with $\Delta\ne0$, the TRM symmetry of the Rice-Mele model still holds: $\hat{\mathcal{S}}H\hat{\mathcal{S}}=H$, where the anti-unitary operator $\hat{\mathcal{S}}=i\sigma_{y}\mathcal{K}m_{x}$ satisfies $\hat{\mathcal{S}}^2=-1$. Here $\mathcal{K}$ is the complex conjugation operator and the mirror symmetry is given by $m_{x}:x\rightarrow-x$. From this symmetry, it is easy to check that if $H\Psi=E\Psi$, then $H(\hat{\mathcal{S}}\Psi)=-E(\hat{\mathcal{S}}\Psi)$, where $\Psi=(\psi_{1},\psi_{2})^{\text{T}}$. Thus the energy spectrum of the bulk is symmetric. Next we will investigate the real-space topology in the present model based on the complete solution to the boundary-value problem of the Rice-Mele model, in the infinite, semi-infinite, and finite domains, respectively.
First of all, we consider the model in the infinite domain $x\in(-\infty,+\infty)$, and then
\begin{equation}
\label{eigenprob}
\begin{split}
(\delta{J}+J\partial_{x})\psi_{2}(x)&=(E-\Delta)\psi_{1}(x),\\
(\delta{J}-J\partial_{x})\psi_{1}(x)&=(E+\Delta)\psi_{2}(x).\\
\end{split}
\end{equation}
Here the two components of wave function $\Psi$ are decoupled in the case of $E\ne\pm\Delta$. The Schr\"{o}dinger equation for each component is $-J^2\partial_{x}^2\psi_{i}=(E^2-a^2)\psi_{i},\,(i=1,2)$, where $a=\sqrt{\Delta^2+\delta{J}^2}$. The general solution is given by
\begin{equation}
\label{infty-bulk}
\begin{split}
\left[\begin{array}{*{20}c}
{\psi_{1}(x)}\\
{\psi_{2}(x)}\\
\end{array}\right]&\sim\left[\begin{array}{*{20}c}
{c_{1}e^{\frac{\gamma}{J}{x}}+c_{2}e^{-\frac{\gamma}{J}{x}}}\\
{\frac{\sqrt{E^2-\Delta^2}}{E+\Delta}(c_{1}e^{\frac{\gamma}{J}{x}-\zeta}+c_{2}e^{-\frac{\gamma}{J}{x}+\zeta})}\\
\end{array}\right],\\
\end{split}
\end{equation}
where $\tanh{\zeta}=\frac{\gamma}{\delta{J}}$ and $\gamma=\sqrt{a^2-E^2}>0$. This state may be interpreted as an in-gap state $|E|<a$ localized at edges ($x=\pm\infty$). Similarly, the bulk states can be obtained by the substitutions: $\gamma{\rightarrow}i\lambda$ and $\zeta{\rightarrow}i\xi$, where $\lambda=\sqrt{E^2-a^2}>0$. One can obtain the other two in-gap states in the case of $E=\pm\Delta$ as follows: for $E=-\Delta$, the solution is
\begin{equation}
\label{infty-edge1}
\begin{split}
\left[\begin{array}{*{20}c}
{\psi_{1}(x)}\\
{\psi_{2}(x)}\\
\end{array}\right]&\sim\left[\begin{array}{*{20}c}
{-c_{1}e^{\frac{\delta{J}}{J}x}}\\
{c_{1}\frac{\Delta}{\delta{J}}e^{\frac{\delta{J}}{J}x}-c_{2}e^{-\frac{\delta{J}}{J}x}}\\
\end{array}\right];\\
\end{split}
\end{equation}
and for $E=\Delta$, the solution is
\begin{equation}
\label{infty-edge2}
\begin{split}
\left[\begin{array}{*{20}c}
{\psi_{1}(x)}\\
{\psi_{2}(x)}\\
\end{array}\right]&\sim\left[\begin{array}{*{20}c}
{c_{1}\frac{\Delta}{\delta{J}}e^{-\frac{\delta{J}}{J}x}+c_{2}e^{\frac{\delta{J}}{J}x}}\\
{c_{1}e^{-\frac{\delta{J}}{J}x}}\\
\end{array}\right].\\
\end{split}
\end{equation}
Note that the above-mentioned in-gap solutions from Eq. (\ref{infty-bulk}) to Eq. (\ref{infty-edge2}) are all blown up at either edge ($x=\pm\infty$), thus the wave functions are indeed non-normalizable in the infinite domain. To make sense of the non-normalizable wave functions, we can at least add one boundary to cut off the wave function and make it normalizable. The non-normalizable in-gap wave function now becomes an edge state adhered to the boundary. Furthermore, the symmetry of the boundary conditions will lead to the symmetry of the energy spectrum of the Hamiltonian.

\begin{figure}
\begin{tabular}{cc}
\includegraphics[width=3.8cm]{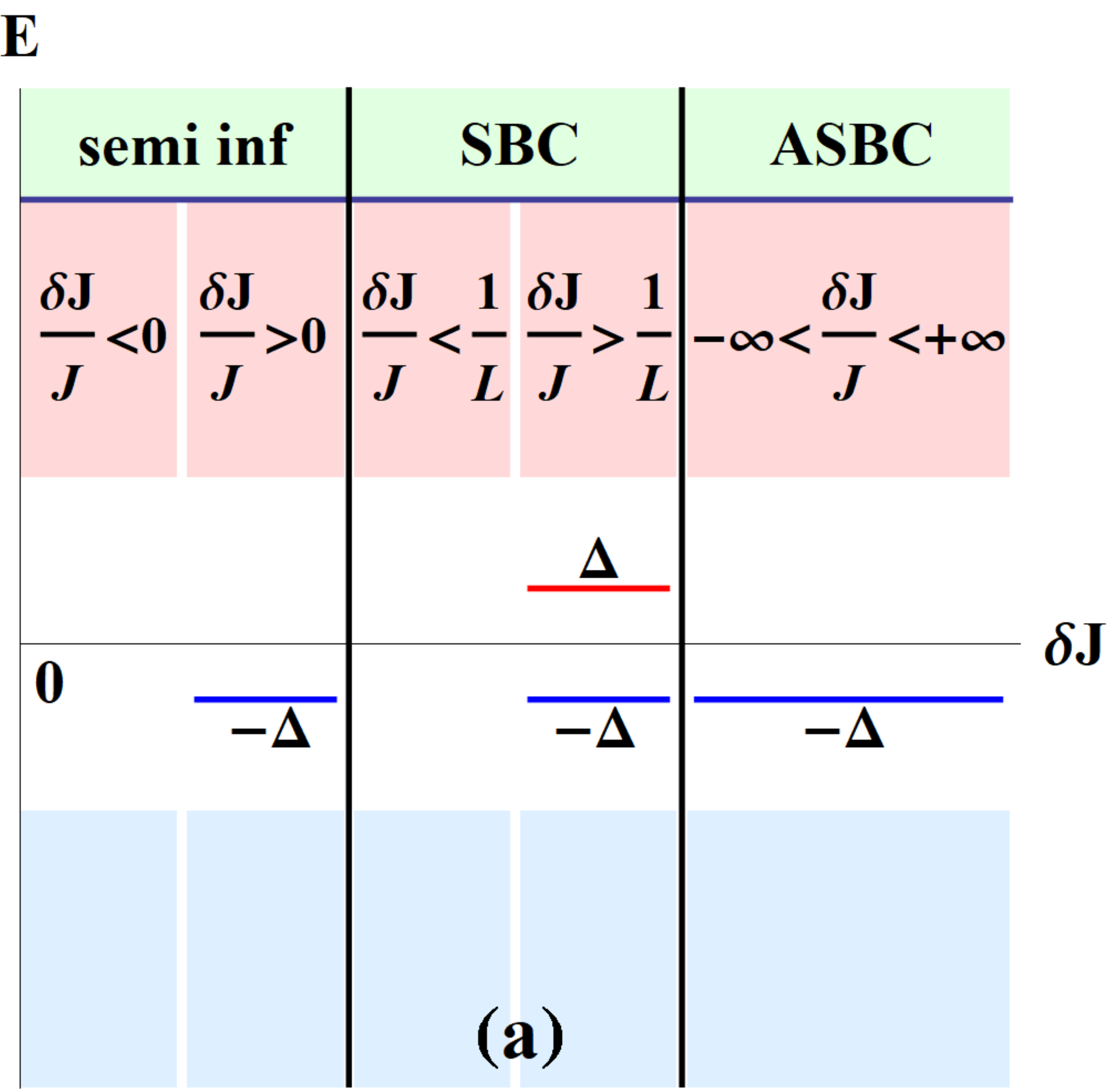} &
\includegraphics[width=4cm]{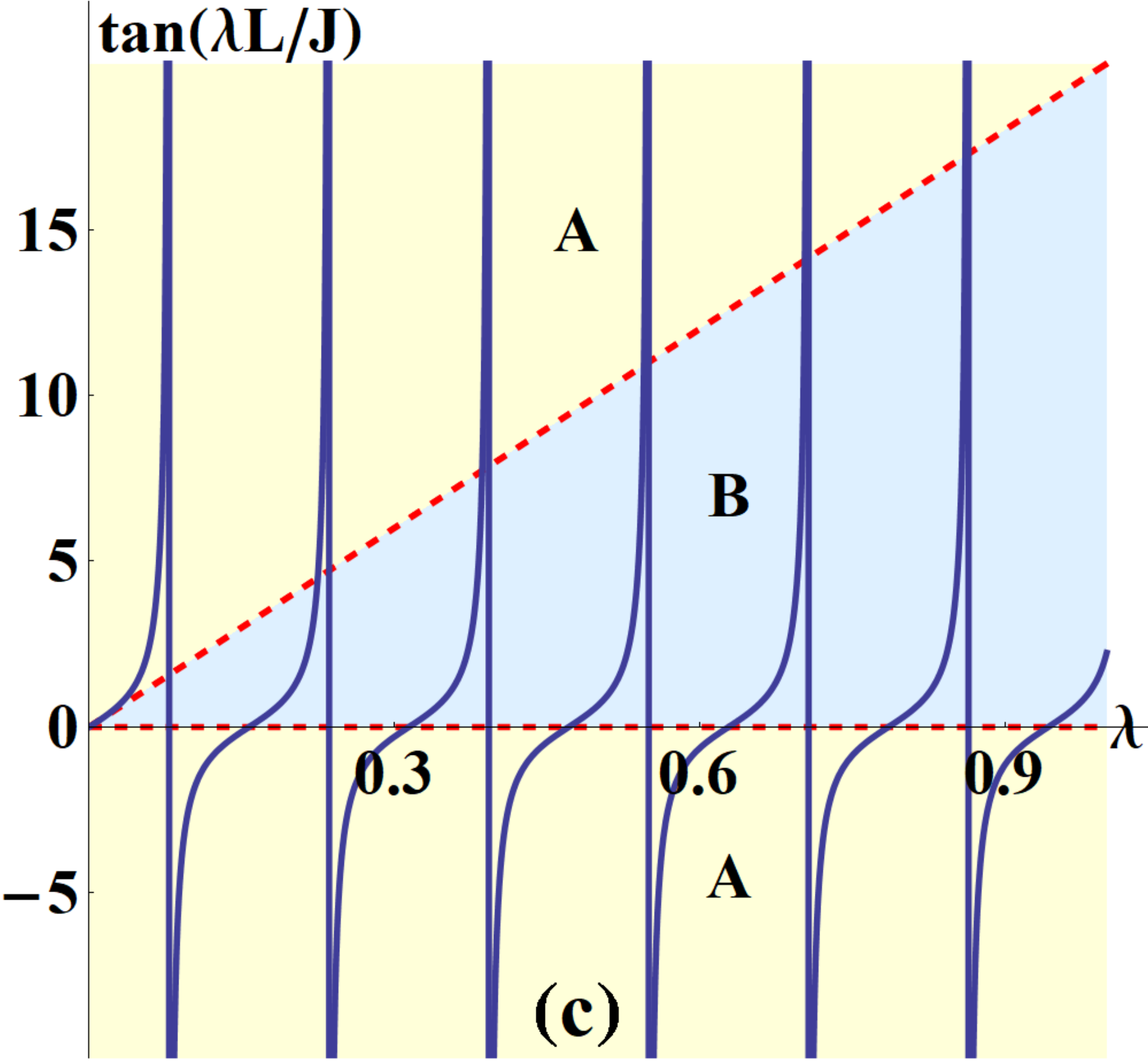}\\
\hspace{-0.4cm}\includegraphics[width=4.4cm,height=4cm]{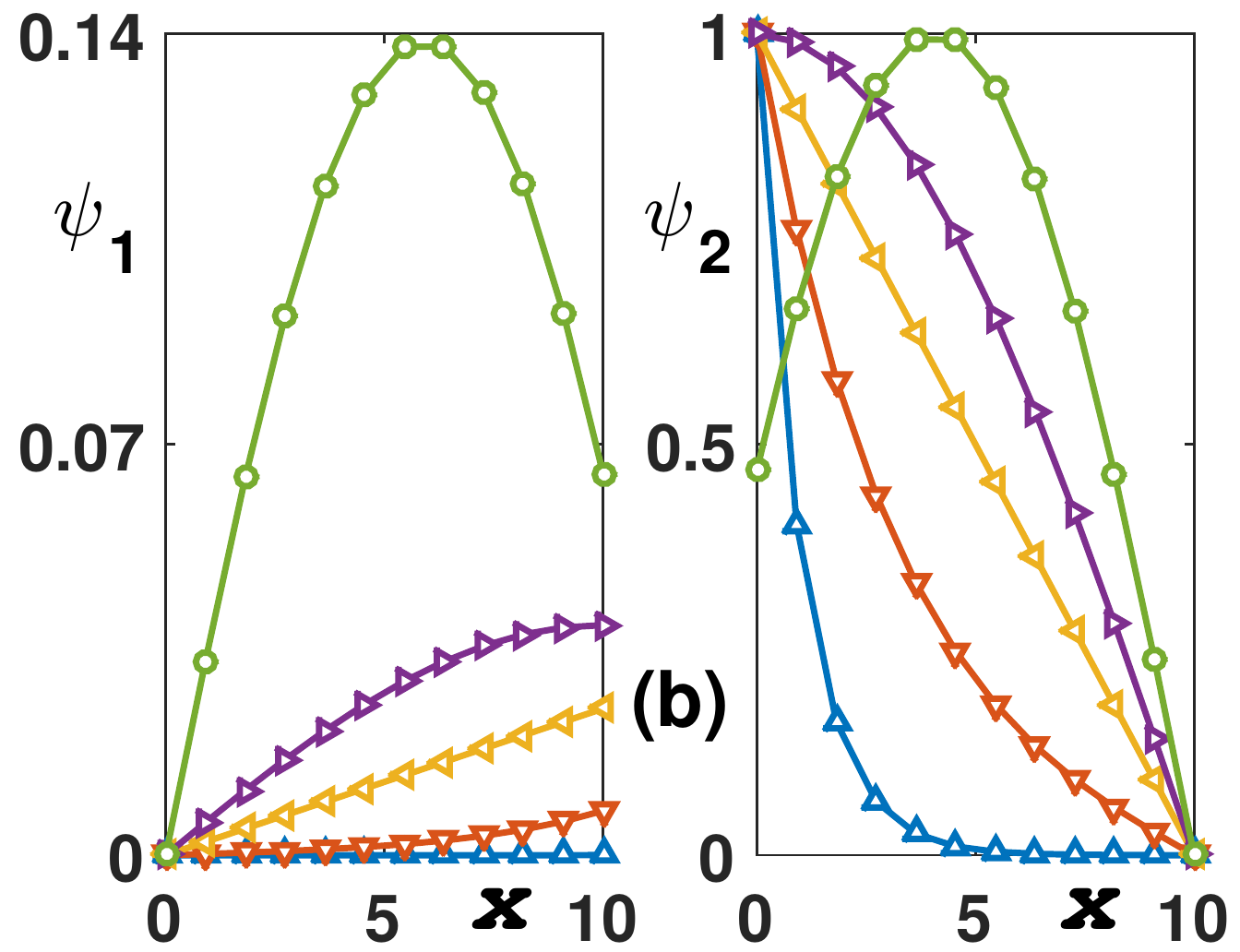} &
\includegraphics[width=4cm]{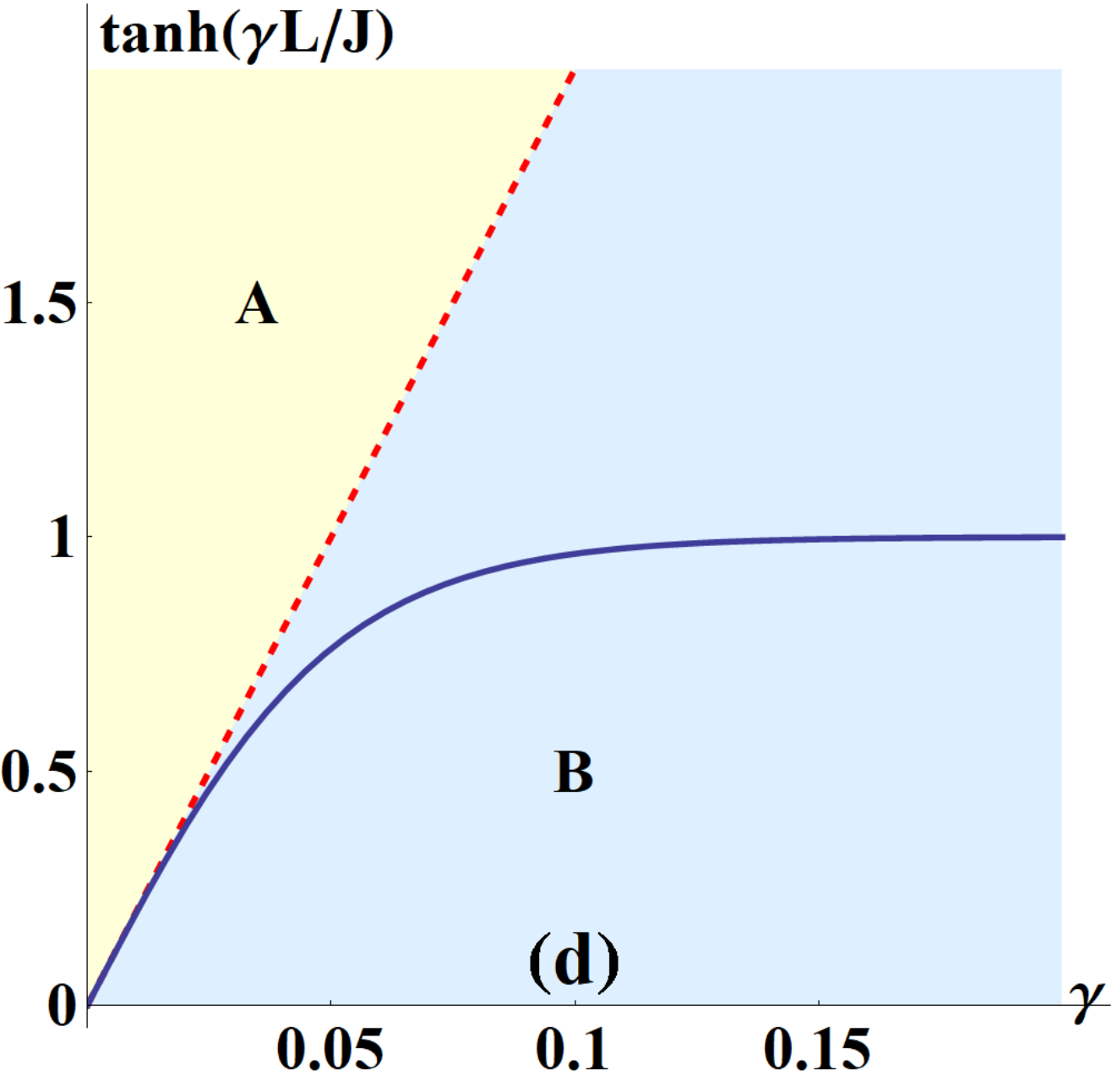}\\
\end{tabular}
\caption{(a) Illustration of the symmetry of the energy spectrum in three different boundary conditions ($J>0$); (b) evolution of the wave function (un-normalized) when $\delta{J}$ crosses the critical point $\delta{J}=\frac{J}{L}$. The parameters are $J=0.5$, $\Delta=1$, $L=10$, and $\delta{J}=-0.25\text{(green,$\circ$)}, 0\text{(purple,$\triangleright$)}, 0.05\text{(yellow,$\triangleleft$)}, 0.15\text{(red,$\triangledown$)}, 0.5\text{(blue,$\triangle$)}$; (c) and (d) are secular equations for Eq. (\ref{finite-edge}). Here region A is in trivial phase while region B is in topological phase.}\label{tantanh}
\end{figure}

Concretely, consider the wave function Eqs. (\ref{infty-bulk})-(\ref{infty-edge2}) in the semi-infinite domain $x\in[0,+\infty)$. To ensure the energy spectrum is real, $\langle\Psi|H\Psi\rangle=\langle H\Psi|\Psi\rangle$, the boundary condition at $x=0$ can be chosen as $\psi_{1}(0)=0$, and the asymptotic behavior satisfies $[\psi_{1}^{*}\psi_{2}-\psi_{1}\psi_{2}^{*}]|_{x\rightarrow+\infty}=0$. From Eq. (\ref{infty-bulk}) we have $c_{1}=-c_{2}$ in the case of $E\ne\pm\Delta$. Thus for $\lambda=\sqrt{E^2-a^2}>0$, the bulk state is a plane wave of the form
\begin{equation}
\begin{split}
\left[\begin{array}{*{20}c}
{\psi_{1}(x)}\\
{\psi_{2}(x)}\\
\end{array}\right]&=\mathcal{N}\left[\begin{array}{*{20}c}
{\sin{\tfrac{\lambda{x}}{J}}}\\
{\frac{\sqrt{E^2-\Delta^2}}{E+\Delta}\sin{(\tfrac{\lambda{x}}{J}-\xi)}}\\
\end{array}\right],\\
\end{split}
\end{equation}
where $\tan{\xi}=\frac{\lambda}{\delta{J}}$ and $\mathcal{N}$ is the normalization factor. Thus the bulk spectrum is symmetric and continuous, $E=\pm\sqrt{\lambda^2+a^2}$ with a full gap $2a$. When $E^2-a^2\le0$, no normalizable solution exists for Eq. (\ref{infty-bulk}). It is found that the in-gap edge state of Eq. (\ref{infty-edge2}) still blows up when $E=\Delta$; while there exists a normalizable edge state for Eq. (\ref{infty-edge1}) when $E=-\Delta$ as well as $\frac{\delta{J}}{J}>0$,
\begin{equation}
\begin{split}
\left[\begin{array}{*{20}c}
{\psi_{1}(x)}\\
{\psi_{2}(x)}\\
\end{array}\right]&=\sqrt{2\tfrac{\delta{J}}{J}}\left[\begin{array}{*{20}c}
{0}\\
{e^{-\frac{\delta{J}}{J}x}}\\
\end{array}\right].\\
\end{split}
\end{equation}
Therefore, the energy spectrum is symmetric when $\frac{\delta{J}}{J}\le0$ but asymmetric when $\frac{\delta{J}}{J}>0$, although the bulk spectrum satisfying TRM symmetry is always symmetric. It implies that a topological phase transition happens when the gap closes at $\delta{J}=\Delta=0$. Here we find that whether $\Delta$ is zero or not, the energy spectrum is qualitatively different when $\tfrac{\delta{J}}{J}$ across 0 as shown in Fig. \ref{tantanh}(a). The spectrum is related to the topology of the Hamiltonian which can be characterized by the fractional Zak phase \cite{Atala2013}.

Now we consider the model in the finite domain $x\in[l_{1},l_{2}]$. Again we impose $[\psi_{1}^{*}\psi_{2}-\psi_{1}\psi_{2}^{*}]|_{l_{1}}^{l_{2}}=0$ to ensure the energy spectrum is real. Accordingly only two types of boundary conditions are available. There are SBC: $\psi_{1}(l_{1})=0,\psi_{2}(l_{2})=0$ and asymmetric boundary condition (ASBC): $\psi_{1}(l_{1})=0,\psi_{1}(l_{2})=0$. For SBC, one can find from Eq. (\ref{eigenprob}) that the boundary condition for $\psi_{1}$ is $\psi_{1}(l_{1})=0,(\delta{J}-J\partial_{x})\psi_{1}(l_{2})=0$ and similar to $\psi_{2}$ when $E\ne\pm\Delta$. This is Robin boundary condition, which is essentially different from the boundary condition of the infinite potential well. Via Eq. (\ref{infty-bulk}), we have $-c_{2}/c_{1}=e^{2i\frac{\lambda}{J}l_{1}}=e^{2i(\frac{\lambda}{J}l_{2}-\xi)}$ and $\xi=\frac{\lambda}{J}L$ where $L=l_{2}-l_{1}$. Thus the bulk state is
\begin{equation}
\label{finite-edge}
\begin{split}
\left[\begin{array}{*{20}c}
{\psi_{1}(x)}\\
{\psi_{2}(x)}\\
\end{array}\right]&=\mathcal{N}\left[\begin{array}{*{20}c}
{\sin{\frac{\lambda}{J}(x-l_{1})}}\\
{\frac{\sqrt{E^2-\Delta^2}}{E+\Delta}\sin{\frac{\lambda}{J}(x-l_{2})}}\\
\end{array}\right],\\
\end{split}
\end{equation}
where $\lambda$ is quantized and given by the solution to $\tan{\frac{\lambda}{J}L}=\frac{\lambda}{\delta{J}}$ ($\lambda\ge0$) as shown in Fig. \ref{tantanh}(c). Similarly, the edge state can be obtained by replacing $\lambda$ by $-i\gamma$ in Eq. (\ref{finite-edge}), where $\gamma$ is given by the solution to $\tanh{\frac{\gamma}{J}L}=\frac{\gamma}{\delta{J}}$ ($\gamma>0$). For this transcendental equation, from Fig. \ref{tantanh}(d), we find that only one nonzero solution $\gamma_{0}$ exists when $\frac{\delta{J}}{J}>\frac{1}{L}$. It is easy to check from Eqs. (\ref{infty-edge1}) and (\ref{infty-edge2}) that there is no nontrivial solution for $E=\pm\Delta$. Therefore, the eigensystem of the Hamiltonian can be described by the quantum number $\lambda_{n}$. When $\frac{\delta{J}}{J}\in[-\infty,\frac{1}{L}]$, $\lambda_{n}\in\{\lambda_{n}|\tan{\frac{\lambda_{n}}{J}L}=\frac{\lambda_{n}}{\delta{J}},n=0,1,2,\cdots\}$; when $\frac{\delta{J}}{J}\in[\frac{1}{L},+\infty]$, $\lambda_{n}\in\{\lambda_{0}=-i\gamma_{0}|\tanh{\frac{\gamma_{0}}{J}L}=\frac{\gamma_{0}}{\delta{J}}\}\cup\{\lambda_{n}|\tan{\frac{\lambda_{n}}{J}L}=\frac{\lambda_{n}}{\delta{J}},n=1,2,\cdots\}$. Notice that each $\lambda_{n}$ corresponds to a pair of wave functions with TRM symmetry, thus the energy spectrum is symmetric for SBC. Consider a series of Hamiltonians with the parameter $\delta{J}\in[\delta{J}^{\text{i}},\delta{J}^{\text{f}}]$. When $\delta{J}^{\text{i}}<\frac{J}{L}<\delta{J}^{\text{f}}$, we can define a map from $t_{n}\in[0,1]$ to $\lambda_{n}$ which connects the Hamiltonian with parameters $\delta{J}^{\text{i}}$ and $\delta{J}^{\text{f}}$:
\begin{equation}
\begin{split}
\lambda_{n}&=\lambda_{n}^{\text{f}}t_{n}+\lambda_{n}^{\text{i}}(1-t_{n}),\quad n=1,2,\cdots,\\
\lambda_{0}&=\Big\{\begin{array}{*{20}c}
{\lambda_{0}^{\text{i}}(1-\frac{t_{0}}{t_{c}}),\quad t_{0}\in[0,t_{c}],}\\
{-i\gamma_{0}^{\text{f}}\frac{t_{0}-t_{c}}{1-t_{c}},\quad t_{0}\in[t_{c},1].}\\
\end{array}\\
\end{split}
\end{equation}
Note that the map is not holomorphic at $t_{c}$ ($\delta{J}=\frac{J}{L}$). However, when $\delta{J}^{\text{i,f}}$ are both greater or lesser than $\frac{J}{L}$, there exists a holomorphic map connecting the Hamiltonian with parameters $\delta{J}^{\text{i}}$ and $\delta{J}^{\text{f}}$. Therefore, for finite size $L$ with SBC, the topological phase transition happens at $\tfrac{\delta{J}}{J}=\tfrac{1}{L}$. Furthermore, we can see that only the state with quantum number $\lambda_{0}$ is topologically nontrivial; all the other states $|\lambda_{n}\rangle,(n=1,2,\cdots)$ are trivial. The evolution of state $|\lambda_{0}\rangle$ across the critical point $\delta{J}=\frac{J}{L}$ is shown in Fig. \ref{tantanh}(b). It is found that the bulk state becomes edge state when the parameter $\delta{J}$ crosses the critical point.

For ASBC, if $E\ne\pm\Delta$, via Eq. (\ref{infty-bulk}), we have $-c_{2}/c_{1}=e^{2i\frac{\lambda}{J}l_{1}}=e^{2i\frac{\lambda}{J}l_{2}}$ and $\lambda$ is quantized as $\lambda_{n}=\frac{\pi{J}}{L}n$, ($n=1,2,\cdots$). Thus when $E^2-a^2>0$, the bulk state is
\begin{equation}
\begin{split}
\left[\begin{array}{*{20}c}
{\psi_{1}(x)}\\
{\psi_{2}(x)}\\
\end{array}\right]&=\mathcal{N}\left[\begin{array}{*{20}c}
{\sin{\frac{\lambda_{n}}{J}(x-l_{1})}}\\
{\frac{\sqrt{E^2-\Delta^2}}{E+\Delta}\sin{[\frac{\lambda_{n}}{J}(x-l_{1})-\xi]}}\\
\end{array}\right],\\
\end{split}
\end{equation}
where $\tan{\xi}=\frac{\lambda_{n}}{\delta{J}}$. When $E^2-a^2\le0$, no nontrivial solution exists. For the edge state, when $E=-\Delta$, there exists one solution,
\begin{equation}
\begin{split}
\left[\begin{array}{*{20}c}
{\psi_{1}(x)}\\
{\psi_{2}(x)}\\
\end{array}\right]&=\sqrt{\frac{2\frac{\delta{J}}{J}}{e^{-2\frac{\delta{J}}{J}l_{1}}-e^{-2\frac{\delta{J}}{J}l_{2}}}}
\left[\begin{array}{*{20}c}
{0}\\
{e^{-\frac{\delta{J}}{J}x}}\\
\end{array}\right].\\
\end{split}
\end{equation}
However, when $E=\Delta$, no nontrivial solution exists. Therefore, the energy spectrum is asymmetric as shown in Fig. \ref{tantanh}(a). For ASBC, we find that there is always an edge state for any $\tfrac{\delta{J}}{J}$, but it does not have topological phase transition. The symmetry of energy spectrum is sensitive to the boundary conditions as shown in Fig. \ref{tantanh}(a). For the finite domain, SBC leads to a symmetric energy spectrum, whereas ASBC leads to an asymmetric energy spectrum. For the semi-infinite domain, the spectrum is symmetric when $\frac{\delta{J}}{J}\le0$ but asymmetric when $\frac{\delta{J}}{J}>0$. For the infinite domain, it is always symmetric.

\section{THERMODYNAMIC PROPERTIES FOR CHARACTERIZING THE TOPOLOGY OF THE MODEL}
\label{thermodynamic_properties}

The emergence of edge states across critical point $\tfrac{\delta{J}}{J}$ is related to the topology of the Rice-Mele model. It is found that this emergence can be reflected by the thermodynamic properties of the Hamiltonian. We first consider the problem in the semi-infinite domain. The energy spectrum has two continuous branches, $E=\pm\sqrt{\lambda^2+a^2}$ ($\lambda>0$) and a discrete level $E=-\Delta$ (when $\frac{\delta{J}}{J}>0$). After some calculations, the density of states (DOS) is
\begin{equation}
\begin{split}
D(E)&=D_{b}(E)+\Theta(\tfrac{\delta{J}}{J})\delta(E+\Delta),\\
\end{split}
\end{equation}
where the DOS of the bulk is $D_{b}(E)=\frac{1}{\sqrt{1-a^2/E^2}}$ and $\Theta(\tfrac{\delta{J}}{J})$ is the Heaviside function. The grand canonical partition function is $\ln\mathcal{Z}=-\sum_{s}\ln{h_{\epsilon_{s}}}$, where $s$ runs over all microstates. The Fermi-Dirac distribution for the electron with energy $\epsilon_{s}$ is $p_{\epsilon_{s}}=\frac{1}{e^{\beta(\epsilon_{s}-\mu)}+1}$ and for the hole is $h_{\epsilon_{s}}=1-p_{\epsilon_{s}}$, where $\beta=\frac{1}{k_{\text{B}}T}$ and $\mu$ is the chemical potential. Since the Fermi sea contains an infinite number of negative energy states, the expectation values of particle number and energy are divergent; however, the fluctuations of them are well-defined quantities. By the standard technique of partition function, the fluctuation of particle number is given by
\begin{equation}
\begin{split}
\langle{(\delta{N}})^2\rangle&=\Big[\int_{-\infty}^{-a}+\int_{a}^{\infty}\Big]{dE}D_{b}(E)p_{E}h_{E}\\
&+\Theta(\tfrac{\delta{J}}{J})p_{-\Delta}h_{-\Delta},\\
\end{split}
\end{equation}
and the fluctuation of energy is
\begin{equation}
\begin{split}
\langle{(\delta{E}})^2\rangle&=\Big[\int_{-\infty}^{-a}+\int_{a}^{\infty}\Big]{dE}D_{b}(E)(E-\mu)^2p_{E}h_{E}\\
&+\Theta(\tfrac{\delta{J}}{J})(-\Delta-\mu)^2p_{-\Delta}h_{-\Delta}.\\
\end{split}
\end{equation}
\begin{figure}
\begin{tabular}{ccc}
\includegraphics[width=4cm]{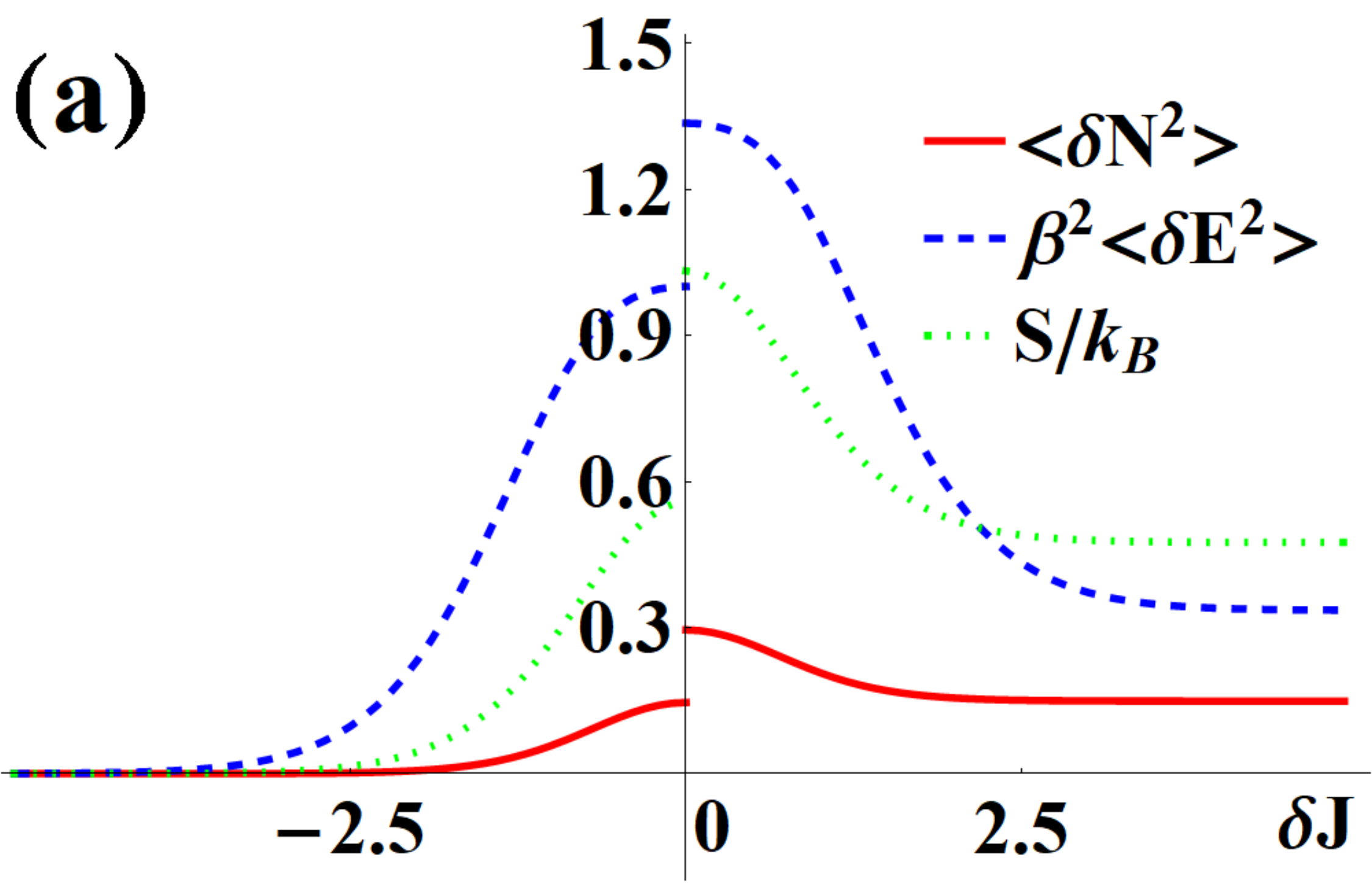} &
\includegraphics[width=4cm]{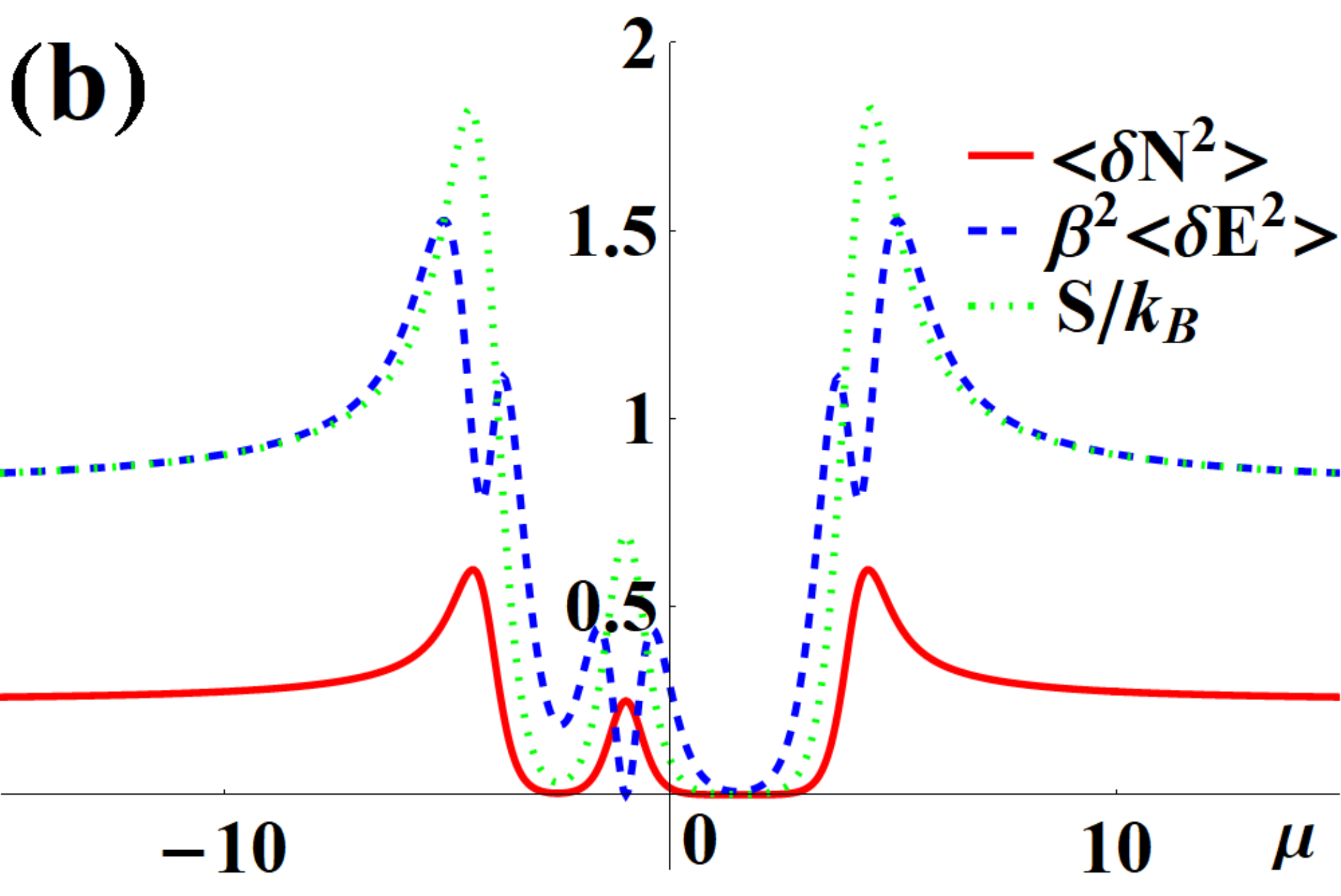}\\
\includegraphics[width=4cm]{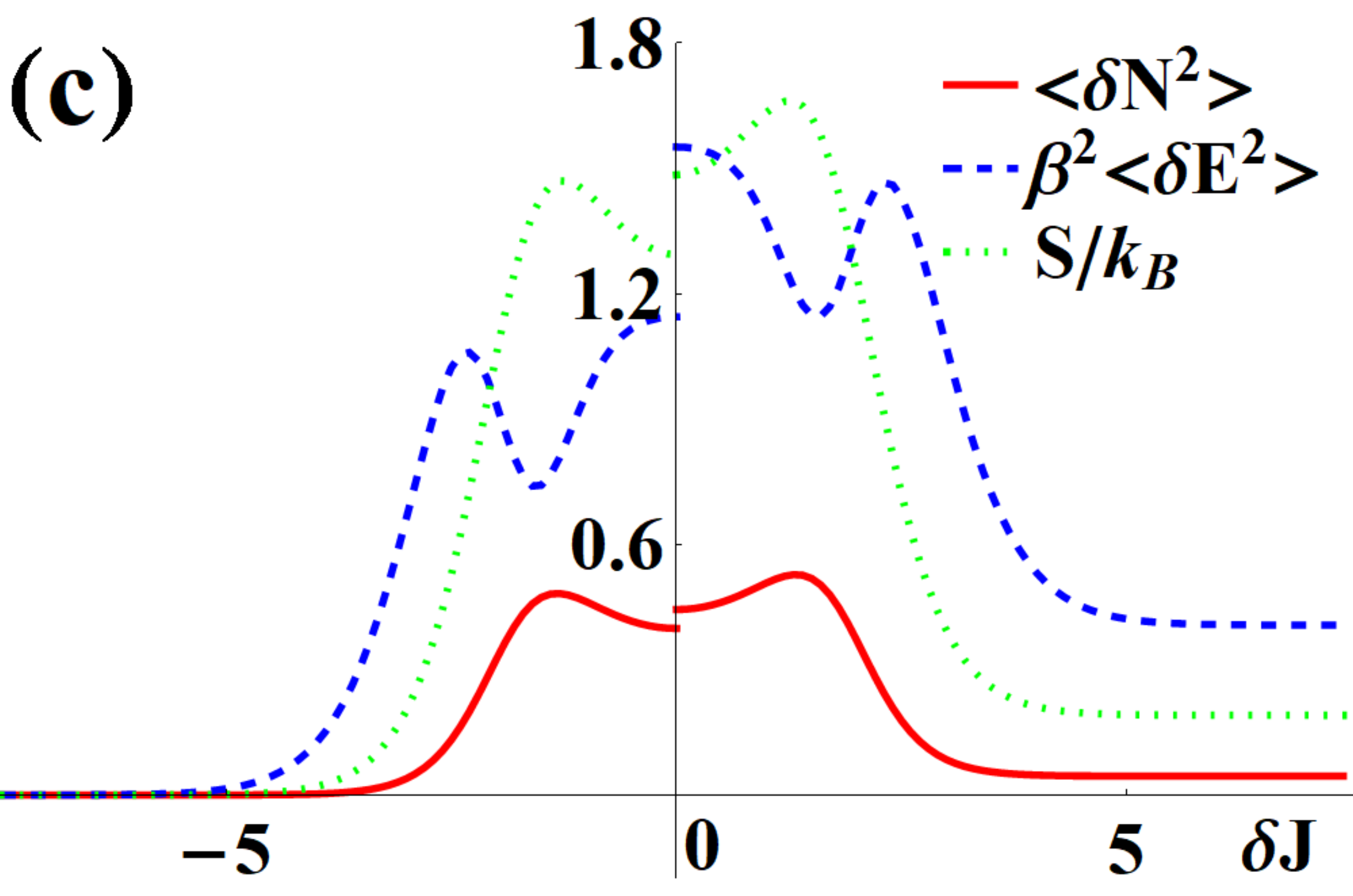} &
\includegraphics[width=4cm]{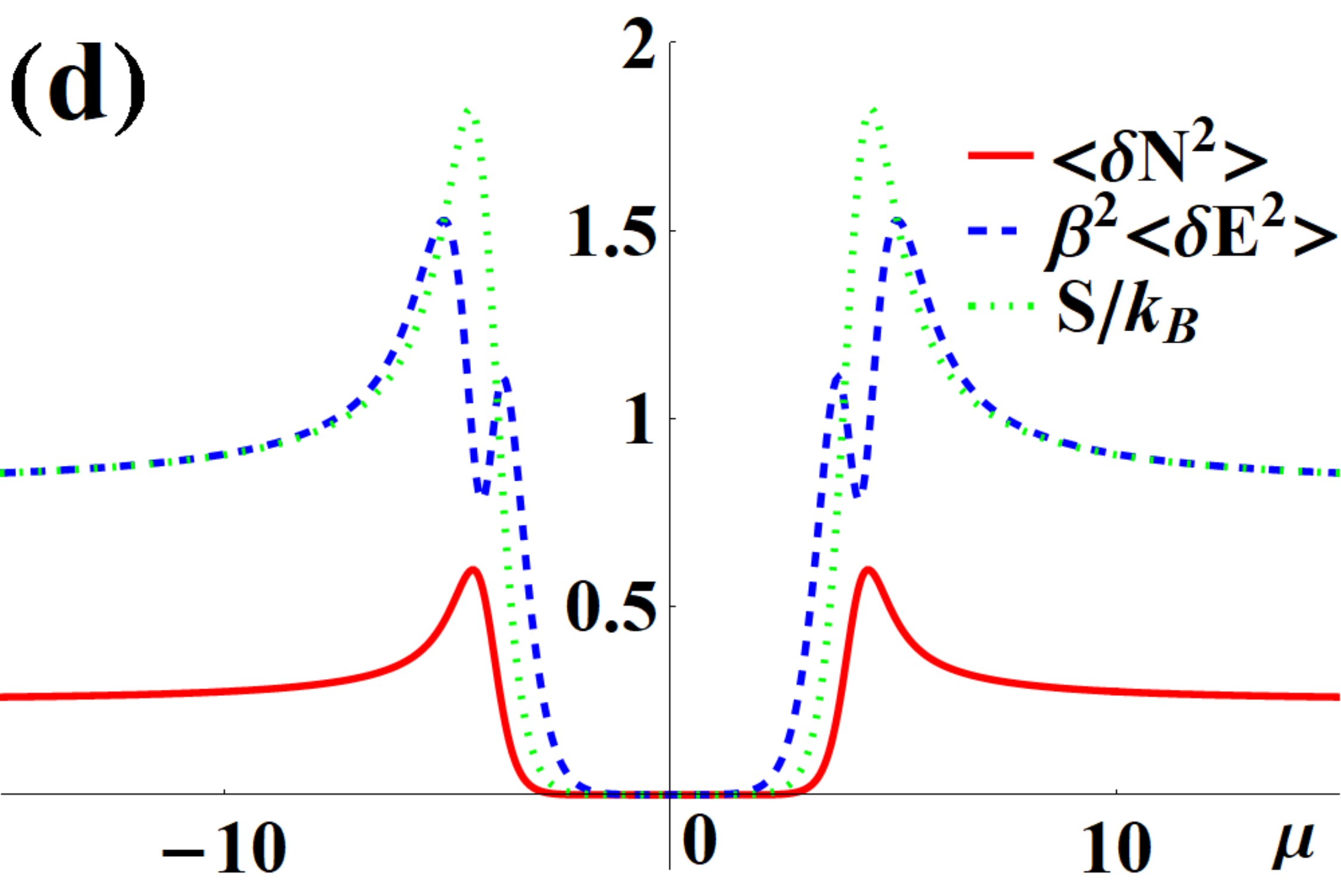}\\
\includegraphics[width=4cm]{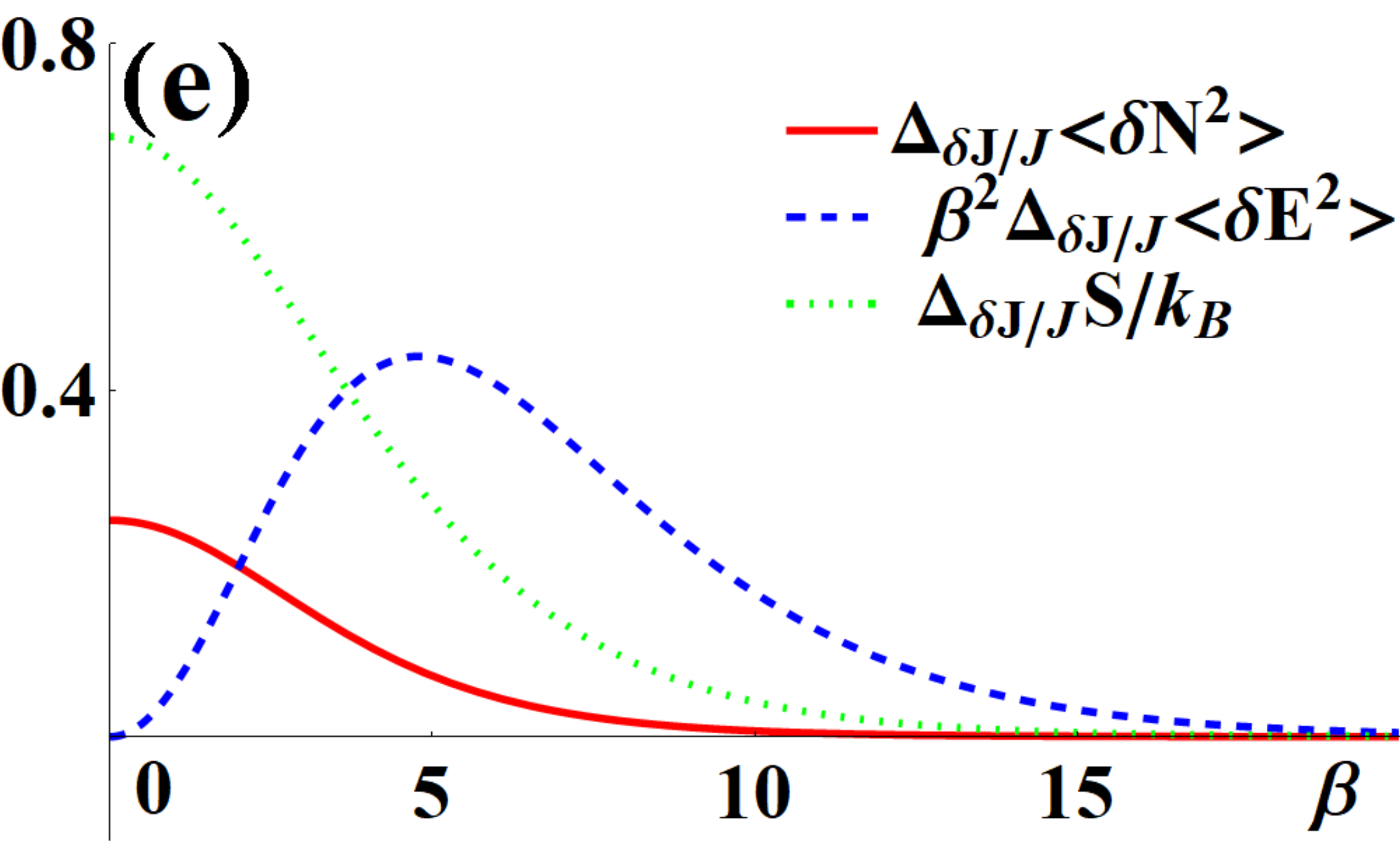} &
\includegraphics[width=4cm]{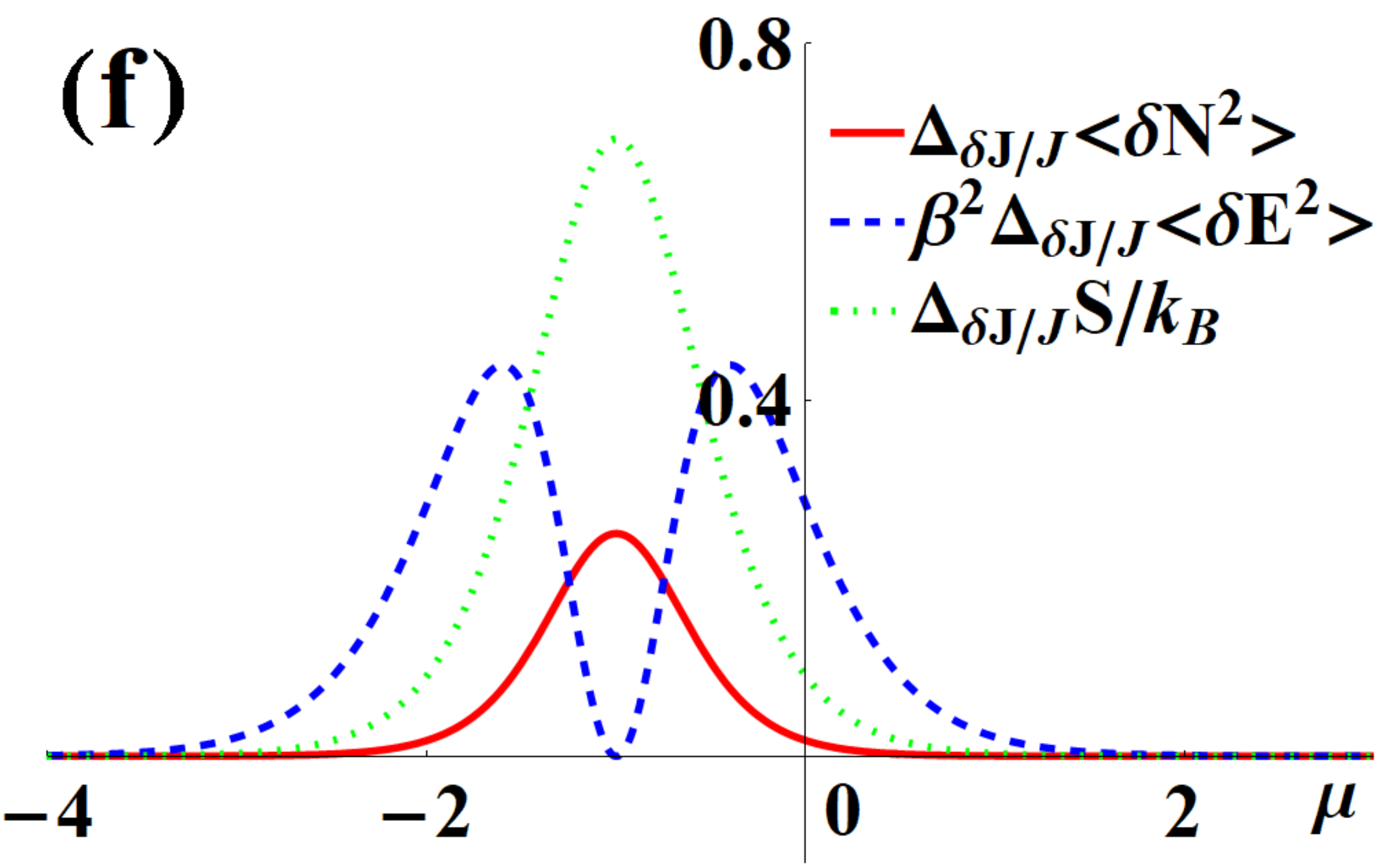}\\
\end{tabular}
\caption{Thermodynamic quantities for semi-infinite domain. The parameters are $J=1$, $\Delta=1$, $\beta=3$ and (a) $\mu=-0.5$; (c) $\mu=-2$. $J=1$, $\Delta=1$, $\beta=4$ and (b) $\delta{J}=4$; (d) $\delta{J}=-4$. (e) $J=1$, $\Delta=1$, and $\mu=-0.5$. (f) $J=1$, $\Delta=1$, and $\beta=4$.}\label{thermoQsemi}
\end{figure}%
Note that the fluctuations of energy and particle number are observable; they relate to the heat capacity and isothermal compressibility by
\begin{equation}
\label{cap_n_compress}
\begin{split}
C=k_{\text{B}}\beta^{2}\langle{(\delta{E}})^2\rangle,\quad\kappa=\tfrac{\beta{L}}{n^2}\langle{(\delta{n}})^2\rangle,\\
\end{split}
\end{equation}
where $n=N/L$ is the particle density of the Rice-Mele chain. The entropy of the system is given by $S=-k_{\text{B}}\beta^2\frac{\partial}{\partial\beta}[\frac{1}{\beta}\ln\mathcal{Z}]$, thus we have
\begin{equation}
\begin{split}
-S/k_{\text{B}}&=\Big[\int_{-\infty}^{-a}+\int_{a}^{\infty}\Big]{dE}D_{b}(E)[p_{E}\ln{p_{E}}+h_{E}\ln{h_{E}}]\\
&+\Theta(\tfrac{\delta{J}}{J})[p_{-\Delta}\ln{p_{-\Delta}}+h_{-\Delta}\ln{h_{-\Delta}}].\\
\end{split}
\end{equation}
The results for the semi-infinite domain are shown in Fig. \ref{thermoQsemi}. Comparing Fig. \ref{thermoQsemi}(b) with Fig. \ref{thermoQsemi}(d), we can see that there is an in-gap peak in the topological phase and the energy spectrum becomes asymmetric for the semi-infinite case. From Figs. \ref{thermoQsemi}(a) and \ref{thermoQsemi}(c), we can find that whether the edge state is occupied or not, there is a discontinuity at the phase transition point $\delta{J}=0$. Furthermore, we observe that the thermodynamic quantities in the limit of $\tfrac{\delta{J}}{J}\rightarrow-\infty$ and $\tfrac{\delta{J}}{J}\rightarrow+\infty$ are different. From Figs. \ref{tantanh}(c) and \ref{tantanh}(d), it is easy to check that the bulk spectra are the same, but there are two more edge states when $\tfrac{\delta{J}}{J}\rightarrow+\infty$. Particularly, by defining the difference as $\Delta_{\delta{J}/J}f=f(\tfrac{\delta{J}}{J}\rightarrow+\infty)-f(\tfrac{\delta{J}}{J}\rightarrow-\infty)$, we have
\begin{equation}
\label{thermoquan}
\begin{split}
&\Delta_{\delta{J}/J}\langle{(\delta{N}})^2\rangle=\sum_{\epsilon}p_{\epsilon}h_{\epsilon},\\
&\Delta_{\delta{J}/J}\langle{(\delta{E}})^2\rangle=\sum_{\epsilon}(\epsilon-\mu)^2p_{\epsilon}h_{\epsilon},\\
&\Delta_{\delta{J}/J}S=-k_{\text{B}}\sum_{\epsilon}(p_{\epsilon}\ln{p_{\epsilon}}+h_{\epsilon}\ln{h_{\epsilon}}),\\
\end{split}
\end{equation}
where $\epsilon=-\Delta$ for the semi-infinite domain and $\epsilon=\pm\sqrt{a^2-\gamma_{0}^2}$ for the finite domain with SBC. Note that these differences are always zero for ASBC. Figures \ref{thermoQsemi}(e) and \ref{thermoQsemi}(f) show the differences as a function of temperature and chemical potential. The differences are always present as long as the temperature is not zero or the chemical potential is near the edge state. They imply the emergence of edge states across the critical point.
\begin{figure}
\begin{tabular}{ccc}
\includegraphics[width=4.2cm]{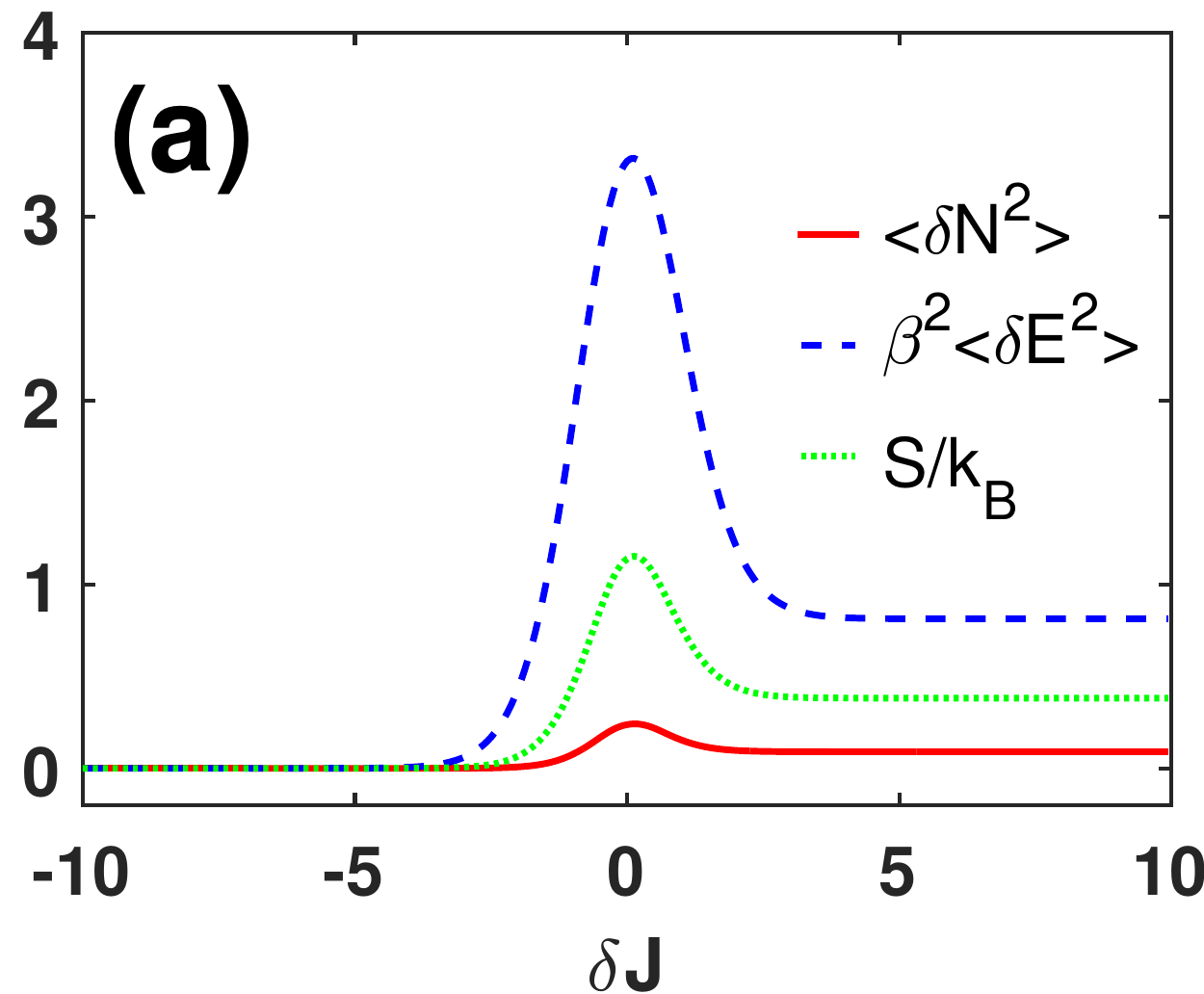} &
\includegraphics[width=4.1cm]{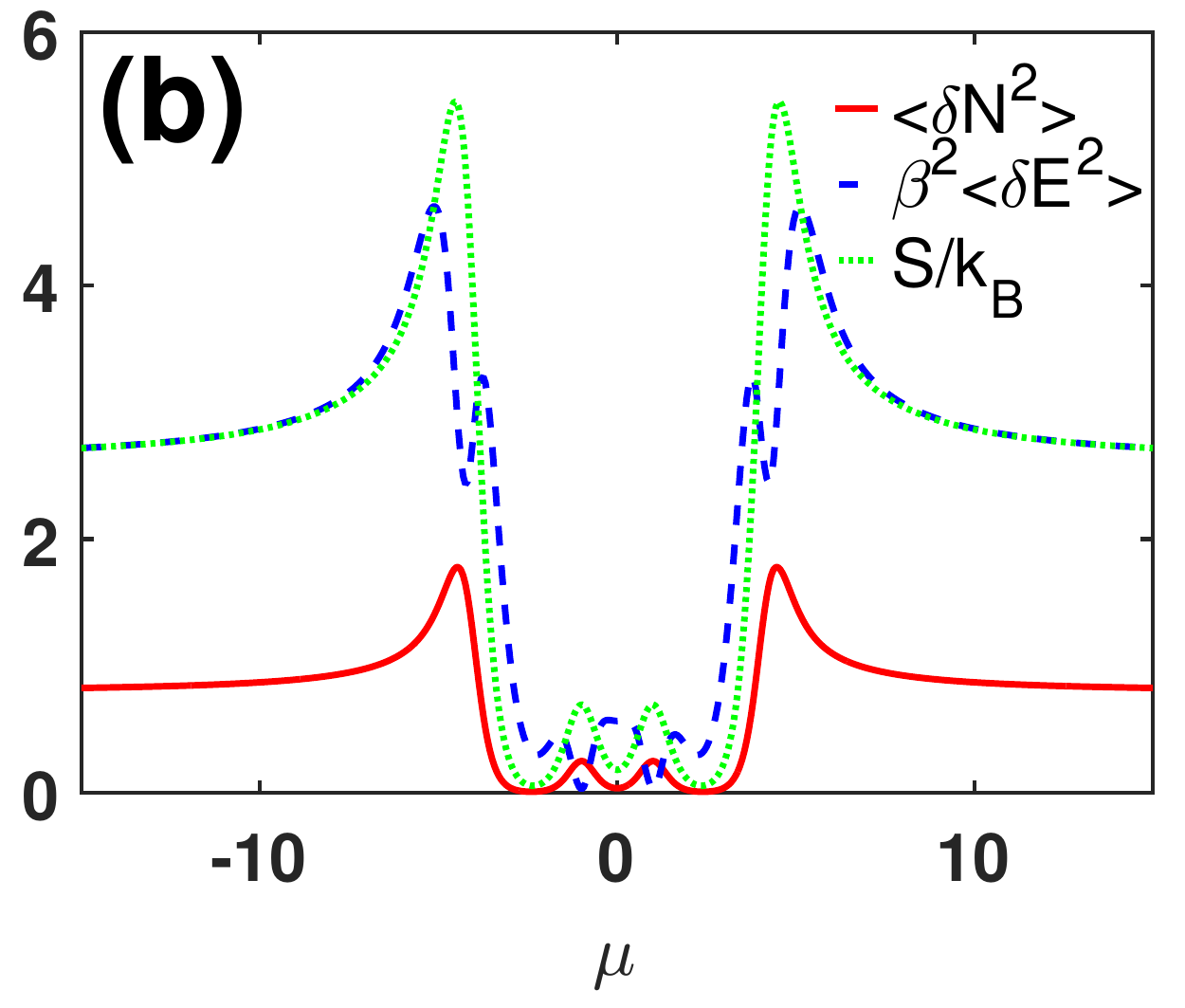}\\
\includegraphics[width=4.2cm]{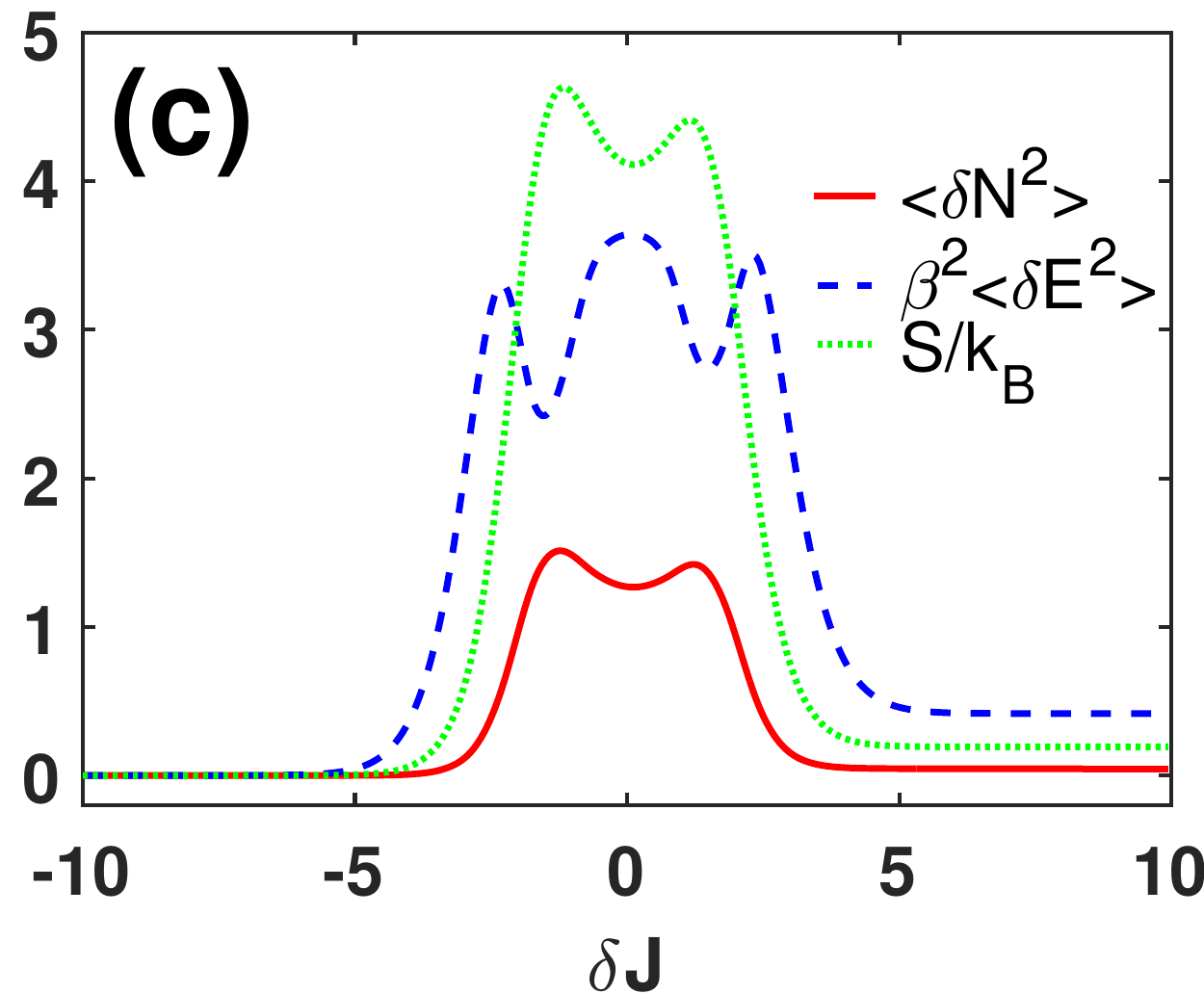} &
\includegraphics[width=4.1cm]{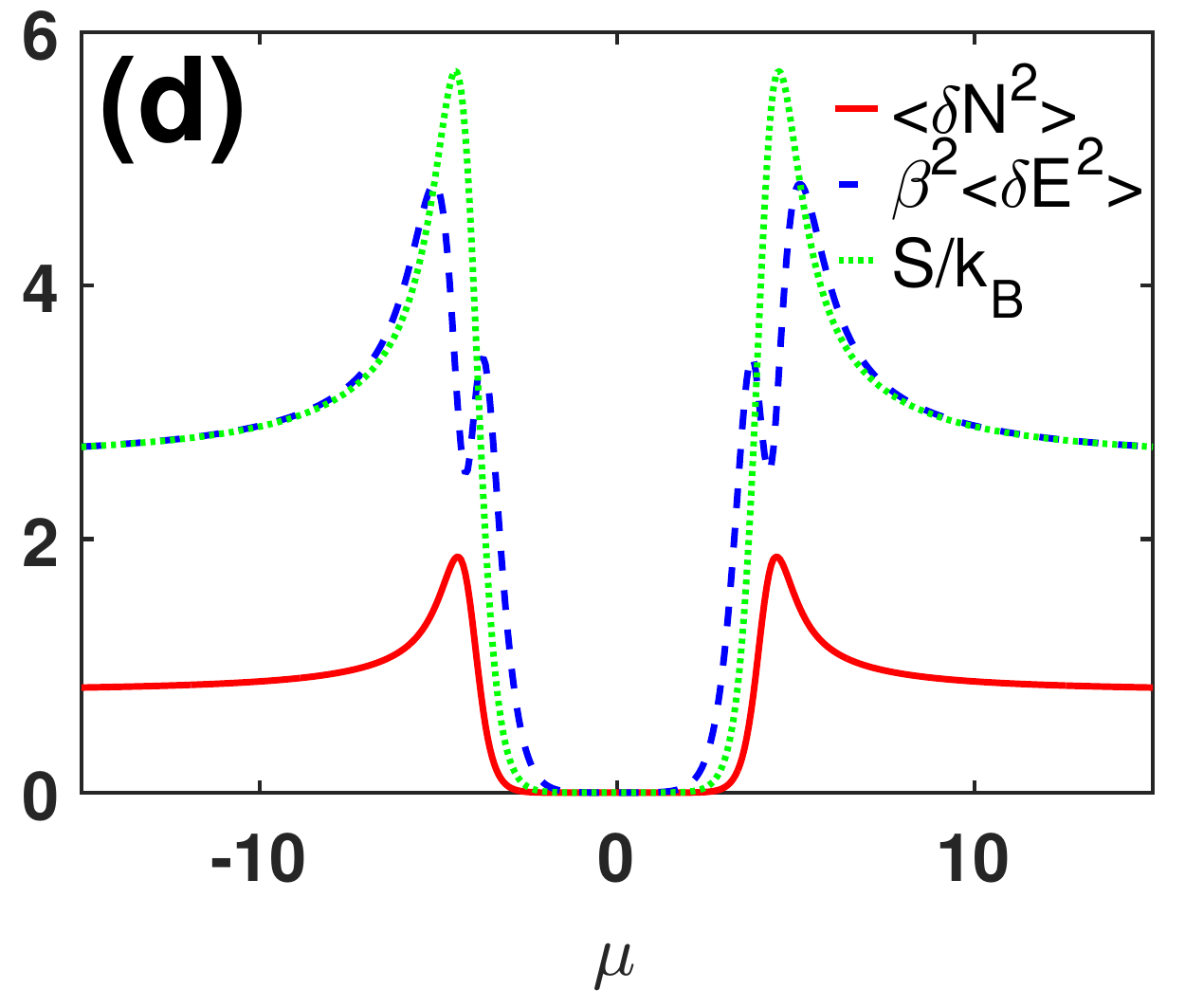}\\
\includegraphics[width=4.2cm]{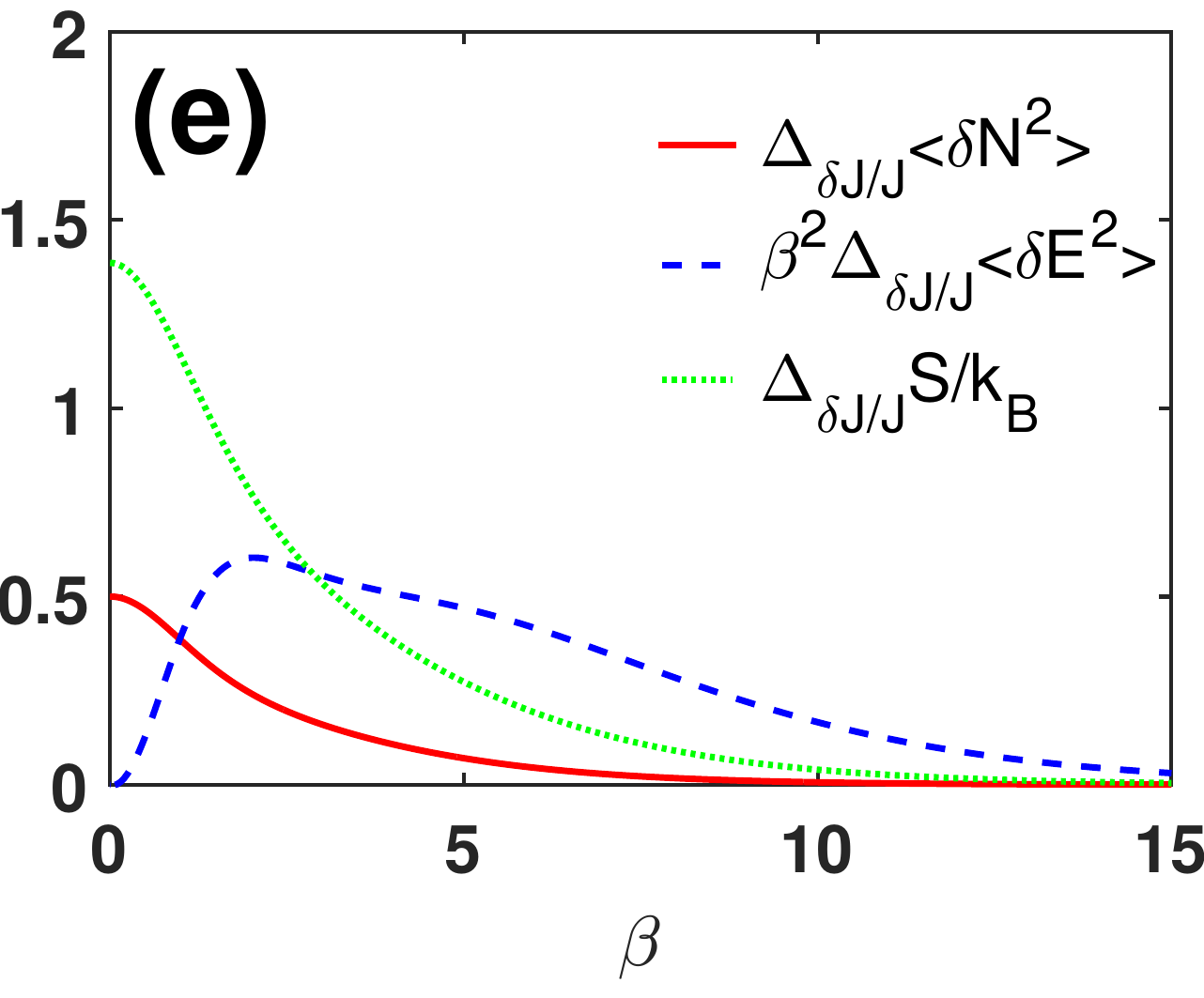} &
\includegraphics[width=4.2cm]{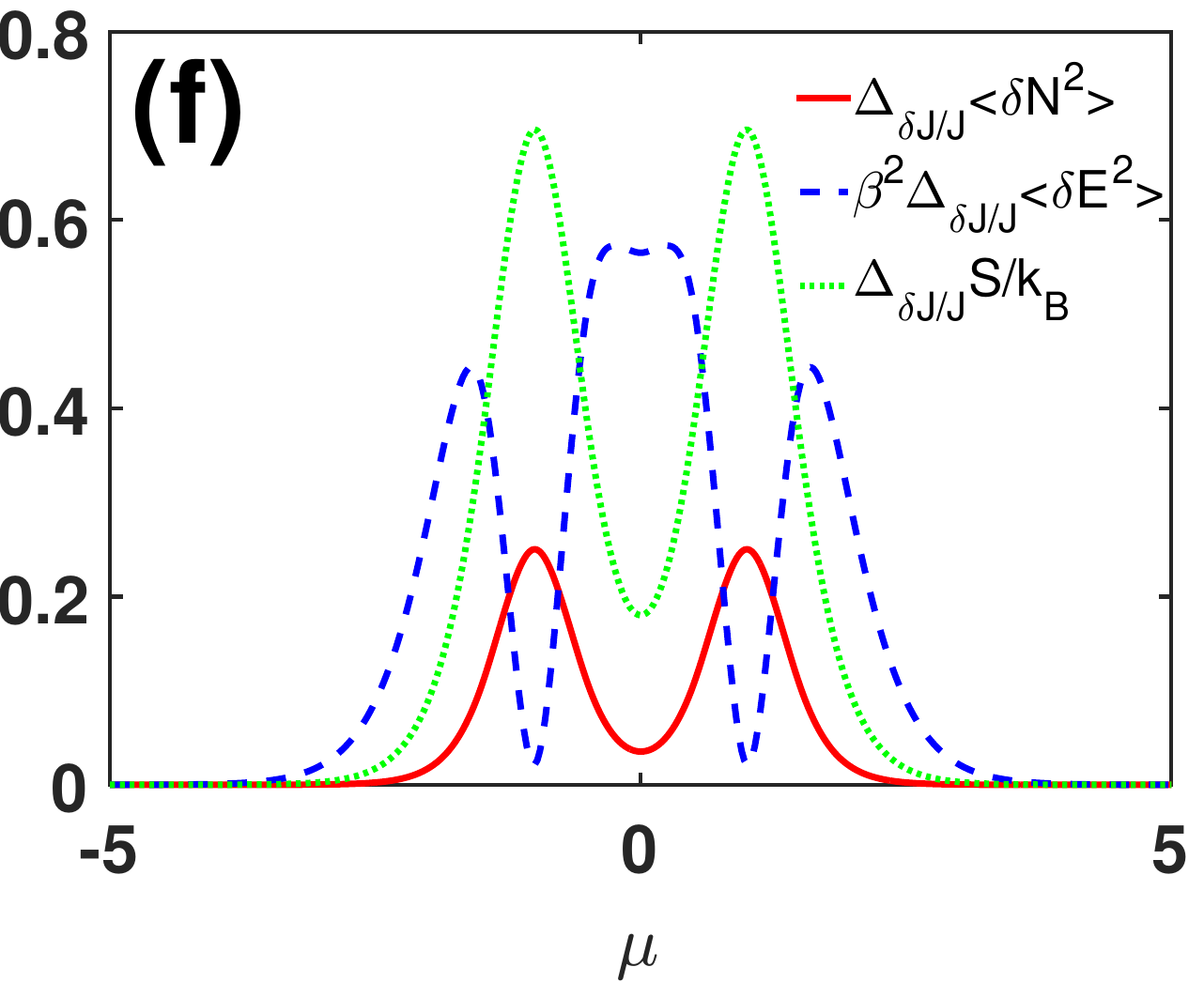}\\
\end{tabular}
\caption{Thermodynamic quantities for finite domain with SBC. We set the size $L=10$. The parameters are $J=1$, $\Delta=1$, $\beta=3$ and (a) $\mu=0$; (c) $\mu=-2$. $J=1$, $\Delta=1$, $\beta=4$ and (b) $\delta{J}=4$; (d) $\delta{J}=-4$. (e) $J=1$, $\delta{J}=15$, $\Delta=1$, and $\mu=-0.5$. (f) $J=1$, $\delta{J}=10$, $\Delta=1$, and $\beta=4$.}\label{thermoQfinite}
\end{figure}
To investigate the thermodynamic properties of the Rice-Mele model in the finite domain with SBC, we can replace the integral $\int{dE}D(E)$ for the semi-infinite domain by $\sum_{n}$, where $n$ runs over all the states in the energy spectrum. From Figs. \ref{thermoQfinite}(a) and \ref{thermoQfinite}(c), we can see that the discontinuity at the critical point shown in Fig. \ref{thermoQsemi} of the semi-infinite case disappears. However, the differences $\Delta_{\delta{J}/J}f$ are still not vanished. This is an indication of the emergence of edge states across the critical point. Similar to the semi-infinite case, the differences always exist when the temperature is not zero. The dependence with respect to temperature is shown in Fig. \ref{thermoQfinite}(e). For the finite domain with SBC, there are two peaks corresponding to the two edge states inside the gap in the topological phase as shown in Figs. \ref{thermoQfinite}(b) and \ref{thermoQfinite}(d). The edge spectrum in the topological phase is symmetric in SBC which is different from the case of semi-infinite domain. The differences $\Delta_{\delta{J}/J}f$ as a function of chemical potential are shown in Fig. \ref{thermoQfinite}(f). The differences are not zero when the chemical potential is near the edge states, but there are two peaks in contrast to the case of semi-infinite domain as shown in Fig. \ref{thermoQsemi}(f).

\section{Experimental realization}
\label{exp_realization}

In the experiment, we could realize the fermionic Rice-Mele Hamiltonian by loading a Bose-Einstein condensate (BEC) of $^{87}\text{Rb}$ into a one-dimensional optical superlattice potential \cite{Atala2013,Nature.448.1029,Nat.Phys.12.296}. The system is then driven into the Tonks-Girardeau limit to mimic the fermionic model by BEC \cite{Nature.429.277,PhysRevLett.92.190401,Atala2013,PhysRevLett.93.210401,PhysRevLett.95.190407}. The superlattice potential is formed by superimposing two optical standing waves of wavelengths $\lambda_{s}=767\text{nm}$ and $\lambda_{l}=2\lambda_{s}=1534\text{nm}$ which are constructed as a stationary lattice (short lattice) and dynamical interferometric lattice (long lattice), respectively. As a result, these laser beams create a lattice potential of the form $V(x)=V_{l}\sin^{2}(k_{l}x+\phi/2)+V_{s}\sin^{2}(2k_{l}x+\pi/2)$, where $k_{l}=2\pi/\lambda_{l}$, $V_{s}$ is the depth of the short lattice, $V_{l}$ is the depth of the long lattice, and $\phi$ is the phase difference between the two lattices whose phase is stabilized and controlled by a Michelson interferometer. Note that $V_{s}$ and $V_{l}$ could be controlled by the respective laser powers and $\phi$ by changing the optical path difference between the two interfering beams with a piezo-transducer-mounted mirror. Phase control between these two standing-wave fields enables us to fully control $\phi$. For example, switching between $\phi=0$ and $\phi=\pi$ allows us to rapidly access the two different dimerized configurations with $\Delta=0$, whereas by tuning $\phi$ slightly away from these symmetry points, we can introduce a controlled energy offset $\Delta$.

Calorimetric studies have long been valuable tools for rigorous tests of physical law \cite{Kinast1296,Ku563,Bloch2012,Ho2010,Zhang1070,Donner1556,PhysRevA.84.013604,PhysRevLett.93.080404,PhysRevLett.77.4984,PhysRevLett.92.120401,PhysRevA.92.063622,PhysRevA.90.043640}, such as the measurements of heat capacity, entropy, and isothermal compressibility of BEC. To measure the heat capacity, we have to transfer a known quantity of energy to the BEC and measure the resulting temperature change.

The energy can be precisely added to the atoms by releasing the cloud from the trap with the influence of gravity and permitting it to expand for a short time $t_{\text{heat}}$ (typically $0\sim1000\mu{s}$), after which the atoms are recaptured and rethermalized \cite{Kinast1296,PhysRevA.92.063622}.
The transferred energy is comprised of three contributions: (i) the atoms fall under gravity and gain kinetic energy; (ii) the displacement $h=\frac{1}{2}gt_{\text{heat}}^{2}$ during fall leads to a potential energy gain when
the trap is reinstated; and (iii) the larger cloud size after the expansion results in greater potential energy when the trap potential is restored. Energy from the first two contributions is $E_{\text{fall}}=N_{\text{Rb}}(\tfrac{1}{2}m\omega_{z}^2h^2+mgh)$, where $\omega_{z}$ is the trap frequency parallel to the direction of gravity \cite{PhysRevA.92.063622}. The expansion energy of the cloud is given by $E_{\text{exp}}=\frac{N_{\text{Rb}}\mu_{\text{TF}}}{7}[2-5\bar{\gamma}^{1.2}+\sum_{i}\gamma_{i}^{2}\lambda_{i}^{2}(t_{\text{heat}})](i=x,y,z)$, where $\gamma_{i}=\omega_{i}(t_{\text{heat}})/\omega_{i}(0)$ is the ratio of trapping frequencies before and after $t_{\text{heat}}$, $\bar{\gamma}=(\gamma_{x}\gamma_{y}\gamma_{z})^{1/3}$, and $\mu_{\text{TF}}$ is the Thomas-Fermi (TF) chemical potential of the initial condensate. Here $\lambda_{i}$ is governed by equation $\ddot{\lambda_{i}}=\frac{\omega_{i}^{2}(0)}{\lambda_{i}\lambda_{x}\lambda_{y}\lambda_{z}}-\omega_{i}^{2}(t)\lambda_{i}$ with $\lambda_{i}(0)=1$ \cite{Blakie2007}. Other methods for transferring energy include an optical phase grating \cite{PhysRevLett.75.4598,PhysRevA.92.063622} or Bragg scattering \cite{PhysRevLett.82.871,PhysRevLett.82.4569,PhysRevLett.86.3930,Blakie2000,Blakie2007}.

On the other hand, the temperature could be measured by time-of-flight imaging with resonant absorption \cite{Kinast1296}. For both the noninteracting and interacting samples, the column density is obtained by absorption imaging of the expanded cloud after $1\sim10\text{ms}$ time of flight, using a two-level states-elective cycling transition \cite{Hara2179,PhysRevLett.92.150402}. The resulting absorption image of the cloud can then be analyzed to determine the temperature of the sample. Thermometry of noninteracting Fermi gas can be simply accomplished by fitting the spatial distribution of the expanded cloud with a TF profile, which is a function of the Fermi radius $R_{\text{F},x}$ and the reduced temperature $T/T_{\text{F}}$ below $0.5T_{\text{F}}$, or of the product $R_{\text{F},x}^{2}{\times}T/T_{\text{F}}$ above $0.5T_{\text{F}}$ where the Maxwell-Boltzmann limit is approached.
Spatial profiles of strongly interacting Fermi gas closely resemble TF distributions,
which were observed experimentally \cite{Hara2179,PhysRevLett.92.120401} and were predicted \cite{PhysRevLett.94.060401}. The profiles of the trapped and released gas are related by hydrodynamic scaling to a good approximation. Similar to the noninteracting case, an experimental dimensionless temperature parameter $\tilde{T}$ can be introduced by fitting the cloud profiles with a TF distribution while holding the Fermi radius of the interacting gas $R'_{\text{F},x}$ constant \cite{Jackson2004}. The temperature calibration is necessary for the above data fitting procedure. We can subject the theoretically derived density profiles \cite{PhysRevLett.94.060401,PhysRevLett.95.260405} to the same one-dimensional TF fitting procedure that was used in the experiments \cite{Kinast1296}.

In addition, the entropy of weakly interacting gas is essentially the entropy of an ideal gas in a harmonic trap which can be calculated in terms of the mean-square axial cloud size $\langle{z^2}\rangle$ \cite{PhysRevLett.98.080402,Luo2009}. For the entropy of the strong interacting gas, we can adiabatically turn up the bias magnetic field until the weakly interacting limit is achieved. Since the process is adiabatic, the entropy during this course is unchanged \cite{PhysRevLett.95.260405}. It is easy to check that the isothermal compressibility in Eq. (\ref{cap_n_compress}) can be recast into $\kappa=\tfrac{1}{n^2}\tfrac{dn}{d\mu}|_{T}$. Since the change in the local chemical potential is given by the negative change in the local potential, $d\mu=-dV$, the compressibility follows as the change of the density $n$ with respect to the local potential $V$ experienced by the trapped gas, $\kappa=-\tfrac{1}{n^2}\tfrac{dn}{dV}|_{T}$ \cite{Ku563,Bloch2012}.

\section{Conclusion}
\label{conclusion}

In conclusion, the thermodynamic quantities have been used  to characterize the real-space topology of the Rice-Mele model. We systematically study the energy spectrum of the model in the infinite, semi-infinite, and finite domains. The non-normalizable wave function for the infinite domain is reduced to the edge state when we add boundaries to the Hamiltonian of the Rice-Mele model. The emergence of this edge state is a signal for the topological phase transition. Furthermore, for the finite domain with SBC, the critical point is $\tfrac{\delta{J}}{J}=\frac{1}{L}$ rather than $0$ as for the semi-infinite domain. In particular, we have studied the model in several different boundary conditions. We find that the symmetry of energy spectrum is sensitive to the symmetry of boundary condition. When the semi-infinite domain or finite domain with ASBC is applied, the edge state is unpaired so that the energy spectrum is asymmetric; whereas when the infinite domain or finite domain with SBC is considered, the energy spectrum is symmetric. The thermodynamic properties which are only related to energy-level statistics can be used to characterize the emergence of the edge state, and subsequently the topological phase transition in the model. We discuss an experimental realization of the Rice-Mele model in the ultracold atom setup and propose the measurements of thermodynamic quantities through the density profile of the condensate.

\section{acknowledgments}
J.B.Y. thanks Feng Mei for fruitful discussions. This work is supported by National Natural Science Foundation of China with Grant No. 11574353 and the National Research Foundation of Singapore under its Competitive Research Programme (Grant No. NRF-CRP 14-2014-04).


%

\end{document}